# Multi-Meron Interactions and Statistics in Two-Dimensional Materials

Xiaobo Lu[1], Linghan Zhu[1], and Li Yang[1,2*]

[1]Department of Physics, Washington University in St. Louis, St. Louis, MO 63130, USA.
[2]Institute of Materials Science and Engineering, Washington University in St. Louis, St. Louis, MO 63130, USA.

**Abstract**

As a fundamental type of topological spin textures in two-dimensional (2D) magnets, a magnetic meron carries half-integer topological charge and forms a pair with its antithesis to keep the stability in materials. However, it is challenging to quantitatively calculate merons and their dynamics by using the widely used continuum model because of the characteristic highly inhomogeneous spin textures. In this work, we develop a discrete method to address the concentrated spin structures around the core of merons. With this approach, we reveal a logarithmic-scale interaction between merons and obtain subsequent statistics of meron gas. The model also predicts how these properties of single and paired merons evolve with magnetic exchange interactions, and the results are in excellent agreement with the Monte Carlo simulations using the parameters of real 2D van der Waals magnetic materials. This discrete approach not only shows equilibrium static statistics of meron systems but also paves the way to explore the dynamic properties of merons through the quantified pairing interactions.

Migrating the concept of soliton from particle physics to condensed matter physics, the topological solitons, such as skyrmions and merons, have attracted tremendous interests in magnetic systems tracing back to several decades ago because of their potential applications for quantum information and storage.[1–6] With substantial experimental advances in recent years, such types of noncolinear spin textures have been observed in various material platforms, e.g., magnetic material surfaces, disks, and thin films within a wide range of temperature.[7–16] The formation of merons merely requires an in-plane O(2) magnetic anisotropy and does not demand any specific type of interactions, such as frustrated symmetric exchange interactions or the antisymmetric Dzyaloshinskii-Moriya (DM) interaction which stabilize the skyrmions.[17,18] Therefore, merons could be one of the common topological solitons in low-dimensional and weak spin-orbit coupling (SOC) magnetic systems.[19–21] More recently, the newly emerging atomically thin two-dimensional (2D) magnets may work as the neat and natural playground for such topological spin textures.[22–25] There are numerous experimental measurements hinting meron states through suppression or large fluctuations of magnetic orders especially for monolayer structures. For example, the suppressed antiferromagnetic (AFM) order of monolayer $NiPS_3$ was reported, and the XY features were discussed in monolayer $CrCl_3$.[26,27]

A systematic theoretical exploration is highly demanded in light of the recent experimental achievements in exploring topological spin textures. The strict 2D planar magnetic vortices and antivortices are known as the BKT physics and have been heavily studied for decades[28,29]. For the topological merons in the natural material, the in-plane geometry of the magnetic vortices and antivortices are inherited. However, unlike the strict 2D model, the out-of-plane spin texture is presented around the core region to lower the free energy of meron in realistic materials. Current studies about meron solitons mainly focus on the structure and stability of bimerons in the frustrated ferromagnetic (FM) or chiral AFM monolayers as well as their dynamical responses under external spin currents or magnetic fields within the Landau-Lifshitz-Gilbert framework.[21,30–34] The accurate description of the structural properties of single meron is limited, which is, however, the foundation for studying bimerons and further collective interactions. To date, most previous works followed the continuum analysis that has been successful in describing skyrmions. For example, the structure of a single meron was mimicked through the micromagnetic simulations (via OOMMF package) accompanied by the continuum analysis[35], and its stability was also discussed through a continuum model based on the Heisenberg Hamiltonian with the DM interaction included[36]. However, different from skyrmions, merons typically exhibit much more compact core structures and form complicated networks in real materials. As pointed out in the seminal work of G. M. Wysin[1] about 2D easy-plane ferromagnets, the region close to the vortex geometry center cannot be described well by a continuum field. This is because the rapid variation of magnetic moments at the core represents a singularity in the continuum view.[1] These characters make the continuum approach questionable to describe general properties of merons. Finally, the statistical distribution of the distance between merons is another determining property of the integrative multi-meron networks in materials, whereas there has been bare rigorous investigation about this fundamental quantity in realistic materials. In this regard, it is worth endeavoring to find a general method to describe the thorough profile of individual merons and their pairing interactions.

In this paper, we develop a discrete approach to study the equilibrium meron properties based on a Heisenberg Hamiltonian that takes the onsite anisotropy and nearest neighbor (NN) exchange interactions into account. By constructing the soliton profile of both single and paired merons based on discrete lattices, we obtain the optimal core size, pair interaction, meron distance distribution and their evolution with the characteristic exchange interaction strength. The model results agree well with the Monte-Carlo (MC) simulations. Thus, this discrete soliton method can be a general way to study merons and provide a comprehensive picture of low temperature meron statistical properties

**Hamiltonian and simulation setup.** We consider the following XXZ-type Heisenberg Hamiltonian with the NN exchange interaction (*J*) and onsite anisotropy (*A*) to describe 2D magnets:

$$\mathcal{H} = A \sum_i (m_i^z)^2 + \frac{J}{2} \sum_{<i,j>} \vec{m}_i \cdot \vec{m}_j \quad (1),$$

where $\vec{m}$ represents the magnetic moment. Our previous study shows that the NN approximation produces a good agreement with the experiment and can predict the characteristic transition temperature of monolayer chromium trihalides within an error bar of several Kelvin.[37] Hexagonal lattices are utilized in this study because most currently synthesized 2D magnets belong to this structure. To mimic realistic materials, we initially set these parameters close to the effective values of monolayer $CrCl_3$[20]: the values of effective onsite anisotropy $A$ the Heisenberg exchange constant $J$ are 35 $\mu eV$ and -790 $\mu eV$, respectively. We have to address that the effective anisotropy $A$ takes into account both the magnetocrystalline anisotropy and the shape anisotropy (magnetic dipolar interactions),[38] and its positive sign results in the in-plane O(2) symmetry that is crucial for realizing merons while prohibiting the formation of long-range magnetic orders at finite temperature. This approximation can capture the essential meron physics while dramatically simplify the model and MC simulations. Utilizing monolayer $CrCl_3$ as starting point, we will keep $A$ fixed while scanning $J$ within a reasonable range to predict the meron properties in general 2D magnets. In other words, we mainly focus on the exchange interaction and the unitless ratio of $J/A$, to avoid scanning a vast parameter space while grasping the essential physics.[39]

Based on the Hamiltonian in Eq (1), we perform MC simulations via the Metropolis algorithm on 2D hexagonal lattices with a size of 240x240 unit cells (if not particularly specified), which contains 115,200 magnetic moments. The periodic boundary condition is implemented in the MC simulations. A MC step consists of an attempt to assign a new random direction in three-dimensional (3D) space to one of the random magnetic moments in the lattice. All magnetic moments are set to point along the out-of-plane direction at the initial state to mimic the experimental cooling condition with the help of an external field.[22,23] We run $2.304 \times 10^{10}$ MC steps (averaged $2 \times 10^5$ steps per magnetic atom) to ensure the equilibrium state is reached.

**Single meron.** We start from the merons in monolayer $CrCl_3$, a 2D magnet expected to be promising to hold in-plane magnetic polarizations and merons.[19–21,27] Figures 1 (a) and (b) present the schematic top and side views of merons obtained from MC simulations of monolayer $CrCl_3$ using the DFT-calculated $J$ and $A$ as mentioned above.[20,37] These figures reveal a few fundamental characters of merons. First, the spin texture has a wave-pocket-like core with the $\pm \hat{z}$ direction spin component, while the easy-plane spin components form the vortex or anti-vortex swirling around the core.[11,14,16,40,41]. Secondly, under ideal conditions, the in-plane swirling of a meron/anti-meron extends to infinity until it is terminated by the system boundary or anti-meron/meron. These features are different from many other widely studied topological spin textures, such as skyrmions or magnetic bubbles.[5,42] For example, a skyrmion can be decomposed into an inner core, an outer domain, and a domain wall separating the core and outer domains as plotted in the inset of Figure 1 (b).[6,38,43]

To quantitatively describe a single meron, we have calculated its topological charge that has been widely employed to characterize the topological properties of spin textures.[5] The topological charge is defined as $Q = \int q(\vec{r}) = \int \frac{1}{4\pi} \vec{n} \cdot \left(\frac{\partial \vec{n}}{\partial x} \times \frac{\partial \vec{n}}{\partial y}\right)$,[44,45] where $\vec{n}$ is the unit vector of local magnetic moment $\vec{m}$. The schematic topological charge distribution $q(\vec{r})$ of a meron in monolayer $CrCl_3$ is shown in Figure 1 (c). In this study, the integrated topological charge of the considered merons and anti-merons belongs to the half integer class ($\pm\frac{1}{2}$). The half integer topological charge supports the idea that merons have to form pairs to stabilize in an integer topology number form, which agrees with previous studies and observations.[11,12,16] Moreover, the most topological charge is condensed around the small core. As shown in Figure 1 (c), the diameter of the 5% isoline of the maximum value of local topological charge is about 4 nm, covering the area only about forty unit cells of $CrCl_3$. This is much smaller than the skyrmions observed in $Fe_{0.5}Co_{0.5}Si$, in which the core size is typically around 90 nm.[9] In addition to the topological charge, we have also calculated the magnetic energy distribution based on the Heisenberg Hamiltonian (Eq. 1). In Figure 1 (d), the magnetic energy profile shows the similar features as the topological charge distribution: most magnetic energy is condensed around the core region within a diameter ~ 7 nm. These results conclude that a main character of merons is their highly inhomogeneous spin texture. The condensed core dominates most their energetic and topological properties of the single meron texture. Interestingly, such highly inhomogeneous spin textures were proposed to carry binary information based on the out-of-plane spin direction of cores and can be manipulated by external field pulse, which is not achievable for the strict 2D planar vortexes.[15] On the other hand, as we will show in the following, such highly inhomogeneous spin textures also bring difficulties in describing and quantifying merons through the traditional continuum analysis.

The first step is to build a modeled profile function to describe the spin texture of a single meron. Then inspired by the hyperbolic secant form of the optic solitons[46], we construct the profile function to describe the meron geometry:

$$\Phi(\phi) = v\phi + \delta \quad (2)$$

$$\Theta(r) = \frac{\pi}{2} - arctan\left[\frac{1}{sinh\left(\frac{r}{w * a_0}\right)}\right] \quad (3)$$

where $\Phi(\phi)$ and $\Theta(r)$ are the azimuthal and polar angles, respectively, which describe the direction of a magnetic moment $\vec{m}$ on the polar coordinate $r(r, \phi)$ (illustrated in Figure 1 (a)), i.e., $\vec{m} = m_0(sin(\Theta)cos(\Phi), sin(\Theta)sin(\Phi), cos(\Theta))$. In Eq. (2), $\delta$ is the phase factor varying from 0 to $2\pi$ describing the swirling. We refer positive vorticity $v$ to a meron and negative $v$ to an anti-meron. In this work, we restrict the discussion to $|v| = 1$, which is the most common and fundamental meron state with $\pm\frac{1}{2}$ topological charge, although $|v| > 1$ can still exist if considering specific exchange interactions beyond NN. [47] In Eq. (3), $a_0$ is the lattice constant of magnetic lattices. $w$ is the scaling size of the meron core region, and it is the only tunable parameter in this model. We can connect $w$ with the widely used full width at half maximum (FWHM) marked in Figure 1 (b). For the soliton form in Eq. (3), FWHM = $2 \ln(2 + \sqrt{3})w \approx 2.63 w$.

Next, we tune the parameter $w$ to get the ground state energy of a single meron soliton based on the Hamiltonian in Eq. (1). One popular way to evaluate such energy is to use the continuum approximation:[2,36,43,48]

$$E = \iint \left(A'm_z^2 + \frac{J'}{2}|\nabla\vec{m}|^2\right) dS \quad (4).$$

$A'$ and $J'$ are treated as the continuous effective coefficients related with Eq. (1). The unitless coefficients ratio $(A'/J')$ is essential, and it takes as $\frac{\sqrt{3}}{3}$ in the hexagonal case. (See section 1 of the Supplemental Information for the detailed processes and the boundary effect) This approach has provided gratifying results for describing skyrmions.[43] In Figure 2 (a), we present the radial profile of merons in monolayer $CrCl_3$ from the continuum model (the dash line) and the MC simulation (the dots). Unfortunately, compared with the MC result, the core size calculated by the continuum model is apparently underestimated. This deviation is from the fact that the main structures of merons, such as the topological charge distribution and the out-of-plane spin texture, are highly concentrated and dramatically varied around the center of the soliton within a few nm close to the singularity point of the continuum treatment. This can be seen from Figure 1 (b), in which sharp variations of the spin texture are observed within a small-sized core.

To overcome such deviation induced by the singularity at the meron core, we can deal with the discrete degrees of freedom of the meron exactly on hexagonal lattices, without the continuum approximations.[1] Following this idea, we discretize the single meron profile function (Eqs (2) and (3)) based on magnetic Bravais lattices and obtain the corresponding energy by directly following the Heisenberg Hamiltonian in Eq (1). Via the procedure of energy optimization, the discretized approach provides a very good portrait for the meron as observed in our low-temperature ($k_B T = J_0/36$) MC simulation of monolayer $CrCl_3$ evidenced by the agreement between the red solid line and dots in Figure 2 (a). Meanwhile, such a good agreement also corroborates the validation of our proposed soliton formula in Eq. (3) for describing merons.

Then we go beyond monolayer $CrCl_3$ to check this discrete approach for general materials. We have calculated the single-meron profile by changing the exchange interaction $J$ within a reasonable range of natural materials, such as chromium trihalides, (from 0.5 $J_0$ to $4J_0$, where $J_0$ is the exchange interaction strength of monolayer $CrCl_3$). The results are summarized in Figure 2 (b). The discrete approach based on the soliton solution (Eq. 3) always provides satisfactory agreements with MC results within such a wide range of magnetic interactions. Moreover, Figure 2 (b) reveals the properties of merons associated with the exchange interaction. For example, a larger exchange interaction will increase the core size of merons. This is similar to the characteristic length scale increase as $\sqrt{J/A}$ in common magnetic domain wall structures. Such a trend is consistent with the understanding that a larger exchange interaction $J$ (in principle, $J/A$) increases the energy cost of forming in-plane swirling at the core region and makes the spins prefer to the $\hat{z}$ direction. Consequently, the soliton favors a broader span of the spin variations and a larger core size. It is worth mentioning that this trend is opposite to overall size shrinkage against the increasing $J$ in skyrmions. When the exchange $J$ increases, the size decrease of the inner region of skyrmions overlays the thickness increase of the domain wall involving the DM interaction, resulting in an overall decreased size of skyrmions.[43]

Figure 2 (c) summarizes the FWHM of merons calculated by the three approaches, i.e., simulated from MC, optimized from the continuum and the discrete approaches, for a wide range of exchange interaction $J$. As explained above, the continuum approach neglects the lattice discretization effect around the core and consequently underestimates the size of the meron soliton. In contrast, by introducing the lattice discretization, the theoretical prediction of meron size is in good agreement with MC simulations. Counter-intuitively, as the $J/A$ ratio increases, the continuous and discrete approaches do not yield the resettling results as observed in typical magnetic domain structures. Such phenomenon is mainly led by the highly concentrated soliton geometry that is kept within a reasonable range of material parameters ($J$ and A).

Finally, it is worth mentioning that the meron core size is relatively robust against the distance between them. Figure 2 (d) displays the MC results of merons' the out-of-plane ($\hat{z}$) spin component under the different pair distances ranging from 10 to 40 nm in monolayer $CrCl_3$. The soliton profiles are nearly the same within the wide range of distance. Notable, the persistence of meron core structure against the pair distance emphasizes the importance of the discrete model in capturing the meticulous structure of the meron core region and verifies the feasibility of studying the interactions among merons through the rigid particle supposition, which will be utilized in the following bimeron profile and multi-meron statistics.

**Bimeron profile.** Unlike a skyrmion which maintains an integer topological charge itself, a single meron only carries half-integer topological charge and shows up in pairs, so-called the bimeron states, to ensure the system belongs to the integer topological class. Hence, their pairing properties are worth being studied. Based on the in-plane geometry of vortex and anti-vortex swirling[11,29], the spin moment ($m(\theta, \phi)$) of a bimeron can be written as Eq. (5) based on the single meron profile illustrated in Figure 3 (a):

$$\begin{cases} \theta = z^+\theta^+(r-r^+) + z^-\theta^-(r-r^-) \\ \phi = \gamma^+\phi^+(r-r^+) + \gamma^-\phi^-(r-r^-) + \delta \end{cases}, \quad (5)$$

where we label the meron and anti-meron via the superscripts by their in-plane swirling type: $\gamma^+ = 1$ for vortex and $\gamma^- = -1$ for anti-vortex, the corresponding vorticities are contained in angles $\phi^{\pm}$. $r^+$ and $r^-$ are the core locations of meron and anti-meron. $z^+$ ($z^-$) indicates the spin direction of the meron (anti-meron) core along the positive (+1) or negative (-1) z-axis. The overall sign of the integrated topological charge of a single meron (anti-meron) is decided by the product of its core region spin direction z and in-plane swirling vorticities γ, which is $N = \frac{1}{2}\gamma z = \pm\frac{1}{2}$.[14,16] For the bimerons discussed in this work, there are three classes of the overall topological index (0 and $\pm 1$). To corroborate results with experiments and MC simulation conveniently, we mainly focus on the class 0 configurations (as illustrated in Figure 1 (b)) where the spin z-direction of the cores in meron and anti-meron are the same. This configuration can be smoothly transferred from the field-induced FM state that could be easily realized in experiments and MC simulations. Since we only consider the isotropic exchange and onsite anisotropy, the phase factor δ can be an arbitrary value between [0,2π) that gives the same degenerated energies. For example, in Figure 3(b) we present the spin structure appearance of meron pairs with four typical values of δ, i.e., $0, \frac{1}{2}\pi, \pi$, and $\frac{3}{2}\pi$, which are observed in the MC simulations.[11,16]

With the above meron pair profile function, we can evaluate the pair interaction energy against the distance ($d = |\vec{r}^+ - \vec{r}^-|$) between meron and anti-meron by using the discrete approach established in single meron profile. First, we begin with merons in monolayer CrCl$_3$. Figure 3 (c) presents the results obtained by the discrete (dots) approach. An attractive interaction is observed between meron and anti-meron. A logarithm curve (dashed line) excellently fits the results for $d > 2r_0$. Interestingly, the Yang-Mills equations, which originally proposed meron-pair quasiparticles in high-energy physics, derives the same logarithm-scale energy variation.[3,4]

Such a logarithm scale can be understood by analyzing the asymptotic behavior far away from the center ($|r| \gg |d|$) via the continuum analysis which is also well established in the BKT physics.[28,29] Due to the highly concentrated meron profile, the corresponding asymptotic interaction between meron and anti-meron mainly comes from the region far from the cores. Thus, it is dominated by the exchange interaction of in-plane swirling of local magnetic moments. To capture the primary scaling of interaction, the energy can be estimated as $E = \frac{J'}{2}\iint \nabla\vec{m} \cdot \nabla\vec{m}\, dS$. Combined with the paired profile (Eq (5)), we could get the asymptotic energy expression as: (The details can be found in Section 2. of the Supplemental Information)

$$E = J'\pi[(\gamma^+ + \gamma^-)^2 \ln\left(\frac{L}{a_0}\right) - 2\gamma^+\gamma^- \ln\left(\frac{d}{a_0}\right)] \quad , (6)$$

where $L$ is the system size. For bimeron states, $\gamma^+ + \gamma^- = 0$ and $\gamma^+\gamma^- = -1$. Therefore, the first term is zero, and the residual second term contributes to a logarithm-scale energy between meron and anti-meron.

We must emphasize that, although the continuum treatment is helpful in understanding the asymptotic behavior, it cannot provide an overall accurate description for the strength of pair interaction. As discussed in the single meron case, the center region of merons cannot be described by the continuum model, and the derivation of Eq. (6) is only asymptotically valid at the large $r$ limit. Therefore, a discrepancy between the continuum model and discrete model should be expected. As shown in Figure 3 (c), the continuum analysis gives an overestimated interaction strength compared with the results from the discrete model, although both exhibit a logarithm scale.

When we study meron interactions, the potential is truncated at $2r_0$, where $r_0$ is the FWHM of the single meron. Such a rigid core treatment for studying medium and long-range interactions at equilibrium state is supported by our MC simulations (Figure 2 (d)) which shows that the meron core properties are barely affected by other merons that are apart larger than $2r_0$. We must clarify that our approach is not able to describe the meron interactions when their cores overlap ($r < 2r_0$), because otherwise the meron profiles no longer follow our assumed soliton formula (Eq. 3). In this sense, we cannot handle the creation or annihilation processes of meron pairs.

To expand our model and discussion to general materials, we have calculated the meron/anti-meron interactions with a range of exchange interaction ($J$), as shown in Figure 3 (d). For all the studied values of exchange interactions, the interaction energy keeps the logarithm scale. Importantly, Figure 3 (d) shows that, as the exchange interaction $J$ increases, the meron/anti-meron attractive interaction is enhanced. Because the total energy is proportional to the exchange interaction as seen from the Hamiltonian in Eq. 1.

Finally, the scaling law of meron pairing energy vs the system size is known to be important as it provides a scale guidance for the experimental realization of meron solitons in the magnetic domains in natural materials.[42,49,50] Unfortunately, the continuum solution confirms the energy convergence of the system as mentioned in Eq. (6) but it cannot provide the asymptotic behavior with varying system size. On the contrary, the discrete approach reveals that the energy of a pair of merons with a fixed distance $d$ converges as $1/L^2 (\sim 1/N)$ to the system size, as shown in Figure 3 (e), and a system size of 240x240 unit cells are utilized in the bimeron study.

**Nearest neighbor (NN) distance of merons in materials.** With the dispersion of interaction energy shown in Figure 3 (d), more general meron behaviors, e.g., the response to the external field and current induced torques[5,30,31,34] can be studied through the subsequent macro soliton effective models. In the following, we focus on the statistical properties of equilibrium meron pairs in materials. Merons in materials can form complicated structures, such as the hierarchic structures from "dipole like" to "quadrupole like" or even more complicated networks.[12,20,30,51] These topological defects networks were also observed in MC simulations of monolayer $CrCl_3$, as shown in Figures 4 (a) and (b). In such complicated network structures, we calculate the NN distance between merons, [51] which is analogous to the distance between meron and anti-meron in a simple pair as discussed in the bimeron case. By examining the hyperbolic secant shape of the solitons, their vortex or anti-vortex in-plane swirling, and integrated half integer topological charge, we can identify the meron type and its center location on the lattice with the help of a k-means clustering algorithm, hence obtain the NN distance[52] In this algorithm, we set the center of the corresponding meron on the hexagonal lattice site. This choice does not affect the result because our simulation shows that the energy variation of different center position is negligible ($\sim 10^{-4} \mu eV$).

After collecting the ensemble of 1,360 distinct MC simulations, we generate the corresponding statistical distribution of the NN distance shown as the histogram in Figures 5 (a)–(d) with different exchange interactions $J$. In the rest part of this article, we will show that, using the bimeron interaction obtained above, we can develop a phenomenological model to capture the statistical NN distance distribution in Figures 5 (a)-(d).

By imitating the NN model in 2D free gas, we start from a probability integration equation for multi-meron systems within the rigid core approximation.[53]

$$f(r) = \left(1 - \int_{2r_0}^{r} f(r')dr'\right)\rho(r), \quad (7)$$

where $f(r)$ represents the probability of finding an antithesis of a meron at the distance $r$. $\rho(r)$ represents the probability of the bimeron having a distance of $r$ with one soliton fixed at the origin.

We further assume that $\rho(r)$ obeys the exponential relation with energy $E(r)$, which can be obtained from the discrete meron pair profile in Figure 3 (d). Thus, $\rho(r)$ can be written as $\frac{N}{Z} 2\pi r \, e^{-\frac{E(r)}{E(2r_0)}}$, in which $E(2r_0)$ is the reference energy, and the normalization factor $Z_0 = \int_{2r_0}^{L} 2\pi r \, e^{-\frac{E(r)}{E(2r_0)}} dr$. N is the total number of merons achieved from MC simulations. By introducing the energy $E(r)$ which can be obtained from the discrete meron pair profile in Figure 3 (d), we get the model describing the meron/anti-meron distance from Eq (7)

$$f(r) = Z_1 r \frac{N}{Z_0} \exp\left(-\frac{E(r)}{E(2r_0)}\right) \exp\left(-2\pi \frac{N}{Z_0} \int_{2r_0}^{r} r' \, e^{-\frac{E(r')}{E(2r_0)}} dr'\right), \quad (8)$$

where $Z_1$ is the normalization factor. As shown in Figures 5 (a-d), this model is in good agreement with MC statistical results, especially under the long-distance condition. On the other hand, limited by the rigid-core assumption and neglecting possible merging process of merons and anti-merons, there are slight discrepancies at the small (short-range) distance. Nevertheless, the general agreement of our model with MC simulations in Figures 5 confirms that the discrete profile model not only works for single meron but also is suitable for studying multi-meron states in materials, which expands the scope of this approach.

We have also studied the impact of the exchange interaction $J$ on the NN distance of merons. An observed trend from Figures 5 (a) - (d) is that the large exchange strength $J$ gives a shorter averaged NN distance and tighter distribution. Figure 5 (e) plots the highest probability distance and the averaged distance vs $J$. When $J$ increases, the most probable distance exhibits a decreasing trend. This is because a larger J results in stronger attractive interactions between meron and anti-meron as we discussed above. This will intrinsically lead to a smaller NN distance.

Finally, we must emphasize that, although the interaction energy between merons is relatively weak (Figure 3 (d)), this attractive interaction is significant in deciding the statistics of merons. In figures 5 (a)-(d), we plot the result of the non-interacting free 2D gas in a dashed curve. Without the weak attracting interaction, the most probable distance is substantially overestimated. For example, for monolayer $CrCl_3$ (Figure 5 (b)), the ideal gas model gives the most probable NN distance at 210Å while that from the discrete model is 116Å. Moreover, the width of the distribution of ideal-gas model is broader, resulting in a larger averaged NN distance of 267Å. The number is much larger than the discrete model (169Å) because of the neglect of attractive interactions. The similar discrepancies are also observed at different exchange interaction *J*, as summarized in Figure 5 (e).

In summary, we built a discrete approach in light of the seminal discussion of planar 2D magnets[1] to explore the merons' general properties covering the single merons, meron pairs, and their low-temperature equilibrium properties. By comparing with the continuum approach and cross-validating with MC simulation results, we demonstrate that such a discrete profile method accurately captures the important characters of meron solitons in 2D magnetic systems. The results confirm the discrete approach as a concrete way to explore topological meron properties and reproduce their interactions accurately that are crucial for understanding experimental measurements, such as the distance (correlation) distribution of dense nanoscale networks of merons and anti-merons in 2D magnets and 3D magnetic surfaces.[51] Furthermore, with the improved explicit interaction from our discrete model, more interesting models can be constructed to study the dynamic properties of merons in real materials. This model would be useful for exploring the meron profile of pinning, dragging, and tuning in presence of moderate magnetic defects or external dynamic perturbations, which are directly connected with experimental realizations and further promising applications.

**Supporting Information:**

Schematic illustration of meron energy and meron anti-meron asymptotical interaction under continuum approach (.docx)

AUTHOR INFORMATION

**Corresponding Author**

*: E-mail: lyang@physics.wustl.edu (L. Y.)

**Author Contributions**

L.Y. supervised the project. X.L performed the calculations and drafted the paper. All authors discussed the results and edited the paper.

**Notes**

The authors declare no competing financial interest.

ACKNOWLEDGMENT:

The work is supported by the Air Force Office of Scientific Research (AFOSR) grant No. FA9550-20-1-0255. This work uses the Extreme Science and Engineering Discovery Environment (XSEDE), which is supported by National Science Foundation (NSF) grant number ACI-1548562. The authors acknowledge the Texas Advanced Computing Center (TACC) at The University of Texas at Austin for providing HPC resources.

**Figures:**

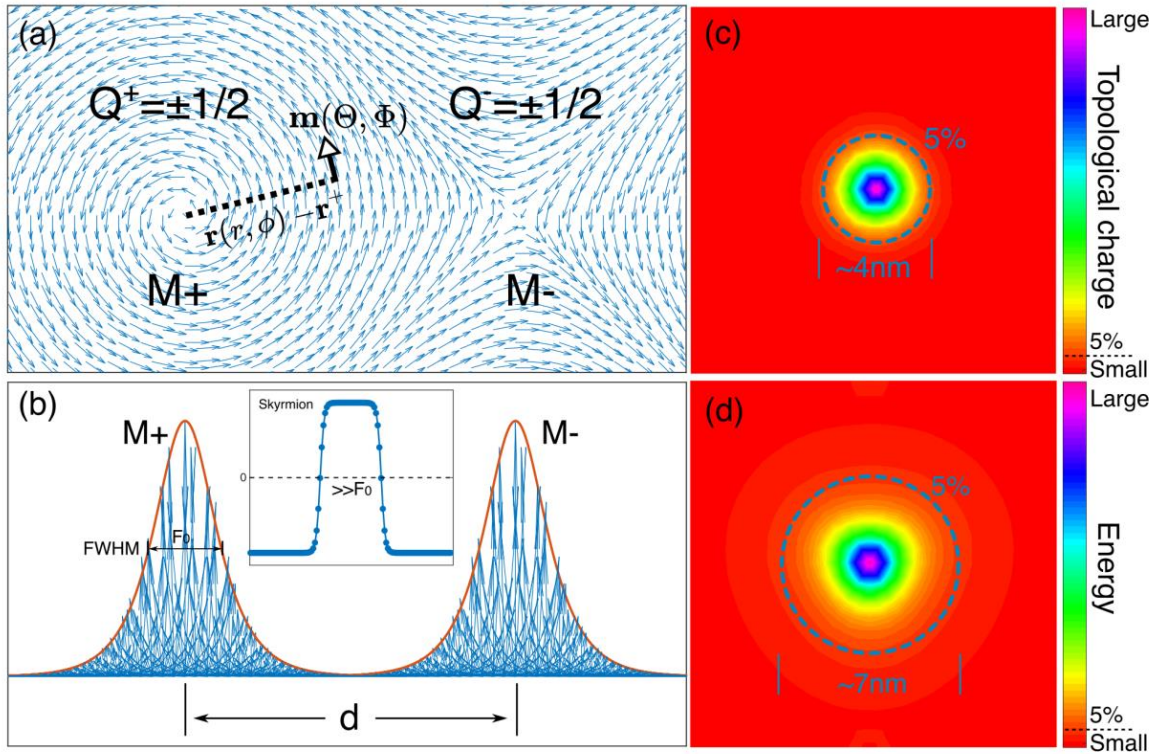

**Figure1** (a) Top view (in-plane chirality) of merons on 2D magnetic lattices. For the articulating purpose, meron and anti-meron are labeled by in-plane swirling types ($M^+$ for the vortex type and $M^-$ for the anti-vortex type). The polar coordinate $\vec{r}(r, \phi)$ and $m$ are illustrated in the dashed line and arrow, respectively (b) Side view of merons. The inset is the structure of a skyrmion. (c) and (d) Schematic figures of the topological charge and energy distribution of a single meron. The dashed circle indicates the isoline with 5% of the maximal value and the estimated diameters are listed in figures.

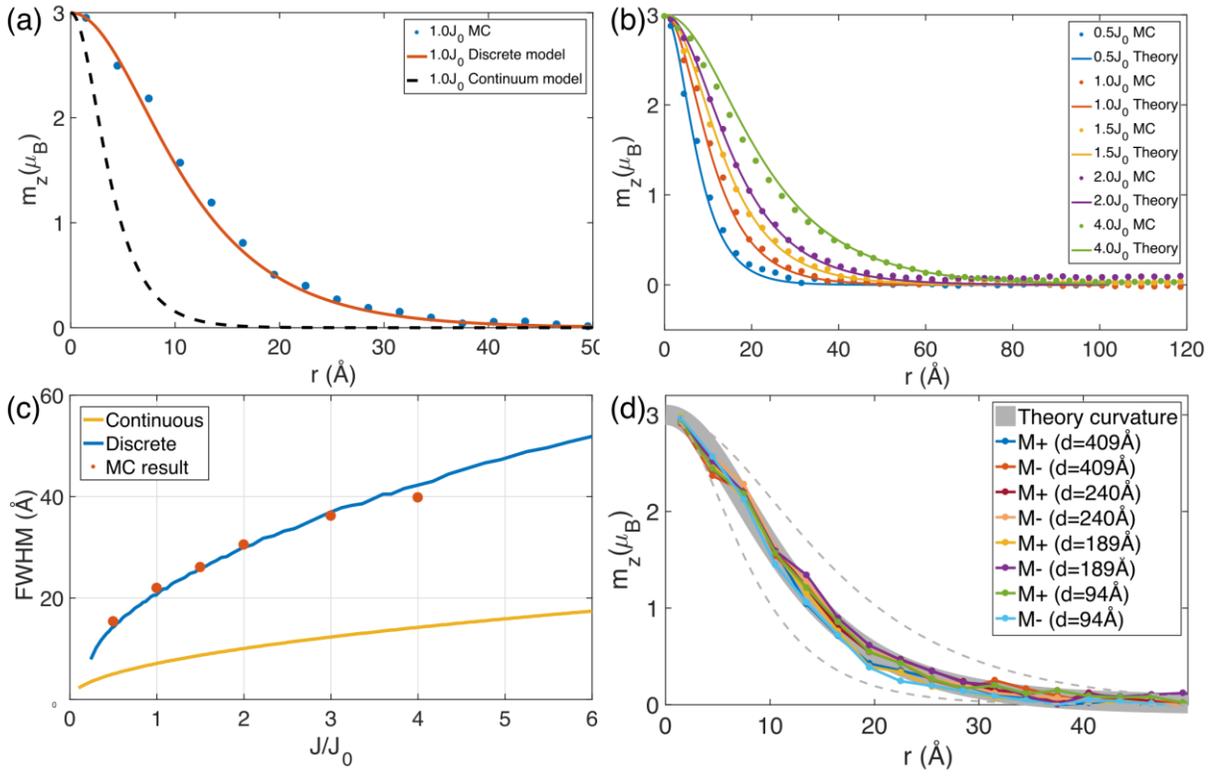

**Figure2** Out of plane spin component ($\hat{z}$) of single meron profile. (a) Dots and lines are the MC simulated results and fitted profile theory results. (b) Comparison of the out of plane spin component ($\hat{z}$) from continuum model, discrete model and MC simulation under fixed $J = J_0$. (c) FWHM of single meron with different $J$ from continuum model, discrete model, and MC simulation. (d) Out-of-plane spin component ($\hat{z}$) of single meron extracted from MC simulation with the meron pair distance under a fixed $J = J_0$. The dashed line indicates the curves of $J = 0.5J_0$ and $J = 2J_0$ from (b) to guide readers' eyes.

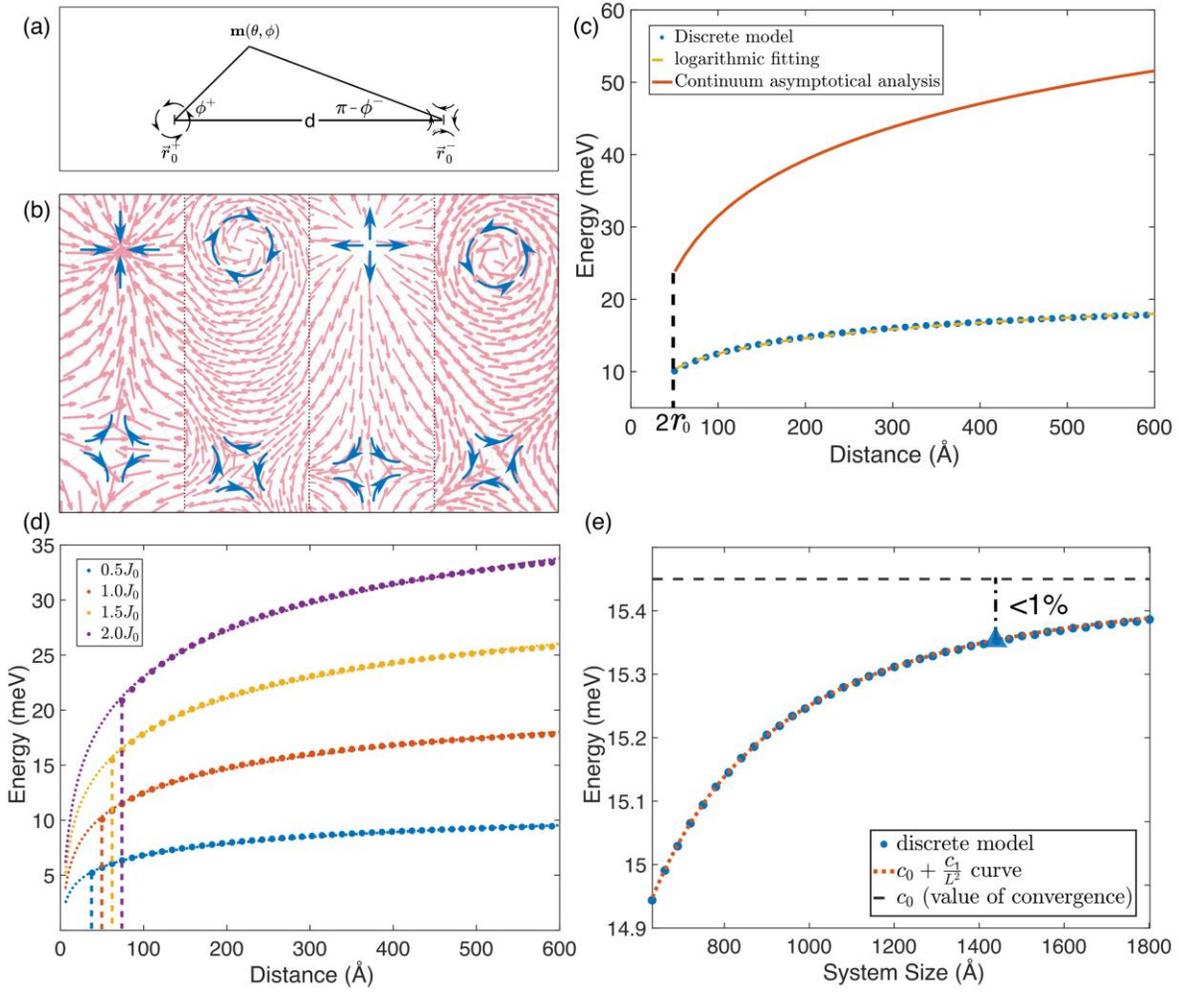

**Figure 3** (a) Schematic of a meron pair and definitions of coordinates (b) Typical meron pairs observed in MC simulations with a phase factor δ close to $0, \frac{\pi}{2}, \pi, \frac{3\pi}{2}$. The in-plane swirling is marked in a 3x3 supercell. (c) Meron pair energy vs the distance between meron and anti-meron with a truncation at $2r_0$. (d) Comparison of pair energies with different exchange interaction $J$. The dot lines are the fitting logarithmic scale. (e) Convergence of meron pair energy against the system size at $J = J_0$ and fixed distance of 240Å.

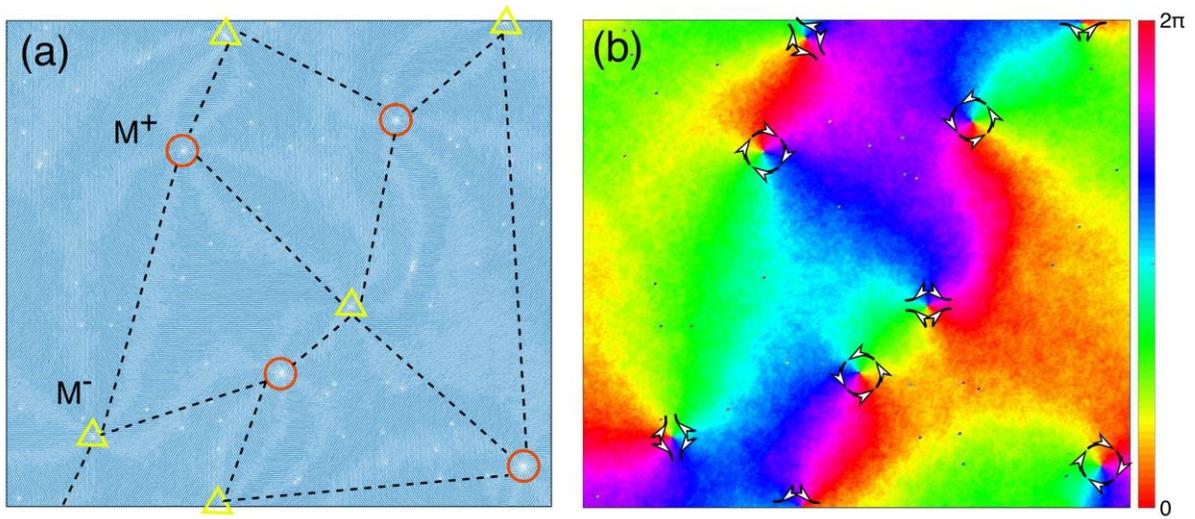

**Figure4** (a) Schematic plot of meron networks from MC simulations of monolayer $CrCl_3$. The merons and anti-merons are labeled by red circle ($M^+$) and yellow triangle ($M^-$) (b) In-plane phase map of (a). The swirling directions of meron and anti-meron are illustrated by arrowheads.

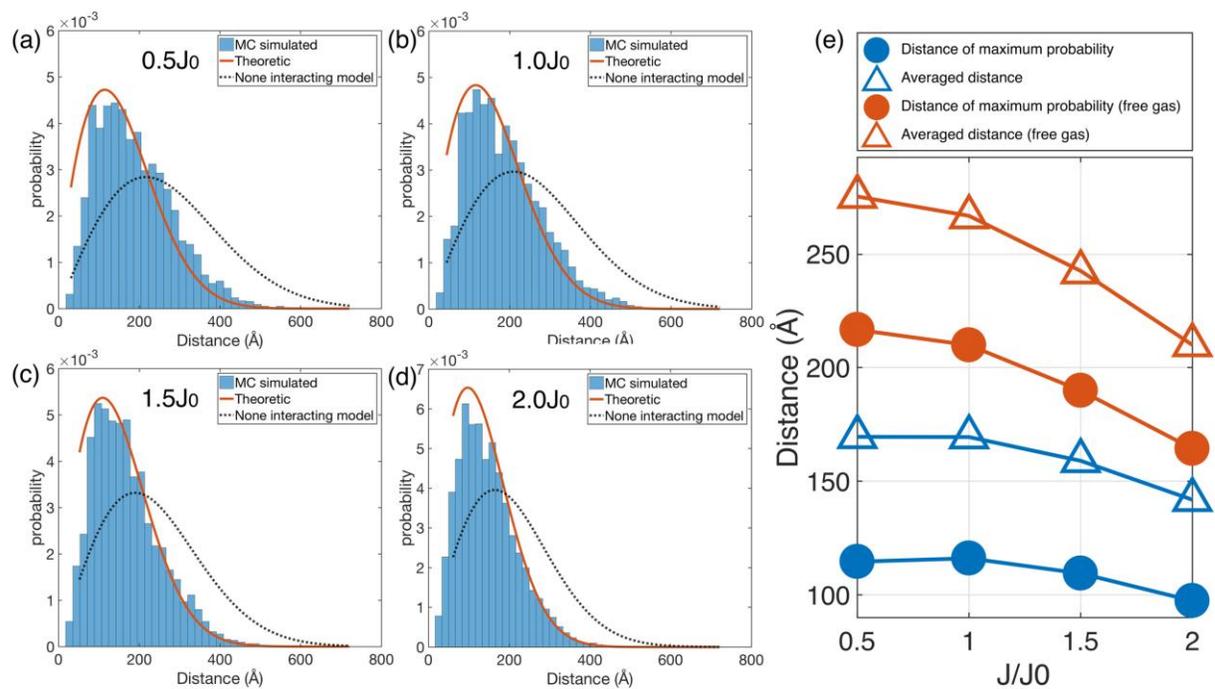

**Figure5** (a-d) Distribution of the NN distance between merons from MC simulations (the blue histogram), the discrete model (the red curve), and none-interacting model (the black dashed curve) under different exchange interaction *J*. (e) Summary of the most probable distance and averaged distance vs the exchange interaction *J*.